\documentclass[12pt,letterpaper,twoside,twocolumn]{article}
\usepackage[latin9]{inputenc}
\usepackage{amsmath}
\usepackage[authoryear]{natbib}

\makeatletter

\pdfpageheight\paperheight
\pdfpagewidth\paperwidth

\usepackage{amsfonts}

\makeatother

\begin{document}
\title{\textbf{Frequentist Inference without Repeated Sampling}}
\author{Paul Vos\thanks{vosp@ecu.edu}\hspace{0.3cm}\&\hspace{0.2cm}Don
Holbert\thanks{holbertd@ecu.edu}\\
 Department of Biostatistics, East Carolina University\\
}

\maketitle
\bigskip{}

\begin{abstract}
Frequentist inference typically is described in terms of hypothetical
repeated sampling but there are advantages to an interpretation that
uses a single random sample. Contemporary examples are given that
indicate probabilities for random phenomena are interpreted as classical
probabilities, and this interpretation is applied to statistical inference
using urn models. Both classical and limiting relative frequency interpretations
can be used to communicate statistical inference, and the effectiveness
of each is discussed. Recent descriptions of $p$-values, confidence
intervals, and power are viewed through the lens of classical probability
based on a single random sample from the population. 
\end{abstract}
\vfill{}

\label{sec:Introduction}

\section{Introduction}

\label{sec:freq_inference} Frequentist inference appears to require
hypothetical repeated sampling. \citeauthor{cox_2006} (2006, page
8) describes frequentist inference as follows
\begin{quote}
Arguments involving probability only via its (hypothetical) long-run
frequency interpretation are called \emph{frequentist}. That is, we
define procedures for assessing evidence that are calibrated by how
they would perform were they used repeatedly. In that sense they do
not differ from other measuring instruments. 
\end{quote}
The entry ``Frequency Interpretation in Probability and Statistical
Inference'' in the\emph{ Encyclopedia of Statistical Sciences }(ESS)
also restricts the interpretation to repeated trials. 
\begin{quote}
$\ldots$ ordinary people ... {[}and{]} many professional people,
both statisticians and physicists, ... will confine themselves to
probabilities only in connection with hypothetically repeated trials.
 \citep{kotz2006encyclopedia}
\end{quote}
\ 

Without proper context these quotes could misrepresent these authors
as only concerned with long-run behavior. \citet{cox_2006} recognizes
the importance of interpreting specific data. 
\begin{quote}
We intend, of course, that this long-run behavior is some assurance
that with our particular data currently under analysis sound conclusions
are drawn. This raises important issues of ensuring, as far as is
feasible, the relevance of the long run to the specific instance. 
\end{quote}
We contend that results from a particular study can be more effectively
described by allowing for a more flexible probability interpretation,
one allowing probability to be interpreted as a limiting relative
frequency or as a simple proportion.

Interpreting probabilities as proportions is the classical interpretation
but has been dismissed because it is viewed as having limited utility.
The entry ``Foundations of Probability'' in the \emph{Encyclopedia
of Biostatistics }states
\begin{quote}
Though influential in the early development of the subject, and still
valuable in calculations, the classical view fails because it is seldom
applicable. \citep{Lindley2005}
\end{quote}
In fact, for understanding $p$-values, in particular, and statistical
inference, in general, the classical view is often applicable. Probabilities
viewed as proportions fit naturally in the context of statistical
inference. Introductory texts use 'frequency' and 'relative frequency'
interchangeably with 'count' and 'proportion', respectively.\footnote{See, for example, \citet{johnson1996statistics} pages 22 and 23. }
In a population, the proportion of individuals having a certain characteristic
provides the same numerical value as the probability that a single
randomly chosen individual will have that characteristic. 

Requiring that frequentist inference include repeated trials is unnecessary
in all, or nearly all, situations. Interpreting probabilities simply
as proportions will allow frequentists to better communicate $p$-values
and other inferential concepts. Furthermore, the classical interpretation
protects against the issue raised by Cox that long-run behavior may
not be relevant to a specific instance. 

\section{Common Understanding of Probability}

To effectively communicate the \emph{p}-value and statistical inference
in general we should know how the term probability, when describing
a random phenomenon, is understood by the general public. Examples
from statistical literature that interpret probability can seem contrived
and do not represent what we observe when considering real world examples. 

\subsection{\emph{ESS} Example}

The following example appears in the aforementioned \emph{ESS }entry\emph{.}
\begin{quote}
A convict with a death sentence hanging over his head may have a chance
of being pardoned. He is to make a choice between white and black
and then draw a ball randomly from an urn containing 999 white balls
and 1 black ball. If the color agrees with his choice he will be pardoned.
\end{quote}
Instead of using the proportion of white balls in the urn to describe
a single random selection, the convict considers an unspecified number
of hypothetical drawings. 
\begin{quote}
The convict replies that he will choose white because $\ldots$ out
of many hypothetical drawings he will in 99.9\% of the trials be pardoned
and in 0.1\% of the trials be executed. $\ldots$ the convict $\ldots$
attaches 99.9\% \emph{probability to the single trial about to be
performed}.
\end{quote}
The article says the convict can attach a probability to a \emph{single}
trial because that 
\begin{quote}
probability is a very real thing to the convict and it is reliably
estimated from past experiences concerning urn drawings. 
\end{quote}
It would seem we need to add the condition that the convict has sufficient
experience with urn drawings. 

Even if that were true, we would expect he would be open to the equally
likely interpretation that clearly applies to a random draw from an
urn. There is no need for a history of ``past experiences concerning
urn drawings'' or a hypothetical future where convicts are executed
repeatedly. 

\subsection{Gambling Examples}

The broadcast of the 2018 Final Table in the 49th No-limit Hold-em
main event held in Las Vegas (aired 13 July 2018 on ESPN's World Series
of Poker) listed the player Cada as having a 14\% chance of winning
while his opponent Miles had an 86\% chance. These probabilities were
based on two cards held by Cada, two held by Miles, and four cards
on the table. These cards were dealt after the deck was thoroughly
shuffled so that each ordering of the 52 cards was equally likely,
or, at least treated as such. There is one more card to be dealt and
the announcer says that Cada has 6 outs \textendash{} cards that would
provide him with a better hand than Miles. There are 44 cards remaining
so the chance that Cada wins is $6/44=$14\%. 

North Carolina, like many states, has a lottery where numbers are
selected by having balls jumbled with shots of air in a confined transparent
space. The Pick-3 game consists of three clear boxes each with 10
balls that are labeled with the numerals 0, 1, ..., 9. These balls
are jumbled for a few seconds and then one is allowed to come to the
top. The jumbling is vigorous enough so that each ball is assumed
to be equally likely to come up. While there may have been some players
who waited for there to be sufficient history of Pick-3 drawings before
placing a bet, we are confident there are many who did not require
such history and still understood the probability of winning.

\subsection{Clinical Trial Example}

The examples above each had a known sample space of equally likely
outcomes and this allowed for the calculation of the proportion that
provided, under suitable randomization, the interpretation for probability.
For statistical inference, simple random sampling from the population
provides equally likely outcomes so that these probabilities can also
be interpreted as proportions. However, unlike the previous examples,
not all population values are known so that proportions cannot be
calculated without specifying a model for these values. 

Consider a trial of 60 participants in which 30 are assigned randomly
to treatment $A$ and the remainder to treatment $B$. For simplicity
we take the response variable to be dichotomous with values favorable
and unfavorable. The population is the 60 participants and the value
for each participant is the ordered pair indicating the outcome, favorable
or unfavorable, under treatment A and under treatment B. Only one
value of each pair is observed. Suppose the number responding favorably
to $A$ is 25 and to $B$ is 17.

One way to compare the treatments is by testing the hypothesis that
the two treatments have the same effect on each participant; that
is, that the values are identical in each of the 60 outcome pairs.
Under this hypothesis there would be exactly 42 favorable responses
regardless of the treatment assignment. The population values consist
of 42 favorable and 18 unfavorable outcomes. By chance 25 of the 42
favorable outcomes were assigned to treatment $A$. Each possible
assignment of 30 outcomes to $A$ can be enumerated and the proportion
where 25 or more are favorable can be calculated. This proportion
is 0.0235. Likewise, the proportion of 25 or more favorable responses
in group $B$ is also 0.0235. The interpretation is as follows: 4.7\%
of all possible treatment assignments have a discrepancy between groups
as great or greater than the observed discrepancy of 25 versus 17.
Because the actual assignment was done in a manner such that each
possible assignment was equally likely, this \emph{proportion} is
the \emph{probability} of an observation as extreme or more extreme
than 25 vs 17. That is, the $p$-value is 0.047 and its interpretation
does not require that we consider hypothetical random assignments
of subjects to treatments. 

\section{Relationship between the Interpretations}

The limiting relative frequency interpretation and the classical interpretation
each describe the same numerical probability. It is not a question
of which is correct. Both are correct and both are available for describing
statistical inferences. The pertinent question is which is more useful
and the answer involves two factors. Before considering these factors
we make a distinction between the definition of probability and an
interpretation thereof.

\subsection{One Definition, Two Interpretations\label{subsec:One-Definition,-Two}}

There generally is wider agreement on how a $p$-value is calculated,
its operational definition, than its interpretation. There are many
incorrect descriptions of the $p$-value but this does not mean there
is only one correct way to interpret its meaning.\footnote{See \citet{Greenland2016} for a useful accounting of misinterpretations.}

The definition of, and confusion surrounding, frequentist inference
involves interpretations of probability, not its definition. Probability
is defined axiomatically as a set function whose domain consists of
subsets from a set $S$, the sample space. When the sample space is
finite the domain can be the power set of $S$. When the sample space
is infinite the power set is replaced with a sigma field. The common
interpretation for the infinite case involves extending the finite
sample space interpretation using limits, in particular limiting relative
frequencies. Another approach is to approximate an infinite model
with one having a finite sample space thereby allowing probability
to be interpreted as a proportion. We follow the latter approach here. 

Epistemologically, what we call an interpretation could be considered
a definition, but our concerns here are more practical than philosophical.
The equally likely definition/interpretation is not intended to cover
every situation where one might use the term probability, but it is
useful for much of statistical inference.  Furthermore, the variety
of settings for statistical inference means proper interpretation
is more easily conveyed when the term probability is not restricted
to only one interpretation. We categorize these settings using two
dichotomous factors: scope and focus. 

\subsection{Scope - Specific or Generic}

The utility of each interpretation will depend on the intended audience.
In the poker example, if the audience is Cada, the player holding
a specific hand, probability is more usefully described as was done
on the broadcast, as a proportion of equally likely cards. More generally,
for casino gambling, if the audience is the house then probability
is usefully described as a limiting relative frequency that describes
an unspecified, but very large, number of hands.

The Lottery example did not include an interpretation of probability.
However, if the audience is a ticket holder, then clearly there is
interest in a specific drawing and the probability is naturally described
as a proportion. On the other hand, the Lottery Commission is more
concerned with on-going drawings and so long-run frequencies are natural
for this audience. 

In the \emph{ESS} example, where the audience is the convict, the
proportion of white balls and the notion of equally likely provide
a simpler description than hypothetical repeated drawings that involve
this or other convicts. The collection of future draws and consequent
executions would be relevant to the state. 

For the investigators of the clinical trial or anyone interested in
the particular outcome of the study, the proportion of randomizations
resulting in a discrepancy as great as 25 and 17 provides a simple
interpretation for the $p$-value. For statisticians interested in
calibrating how inference procedures such as Fisher's exact test ``would
perform were they used repeatedly'' then significance levels would
be specified and probabilities would be described in terms of limiting
relative frequencies of hypothetical repeated randomizations.

The common factor in comparing the potential audience in each of these
examples is the scope, either specific or generic, to which the probability
extends. For a specific outcome, be it a hand of cards that could
determine whether a player continues in the tournament, a lottery
draw for a ticket holder, a convict whose life depends on a single
draw from an urn, or a physician wanting to assess the evidence from
a single study for the merits of a specific treatment, a proportion
provides the natural interpretation for the probability related to
a single randomization. 

The scope is generic when a specific outcome is viewed as part of
a collection and probability describes this collection. For statisticians
who are concerned with how their methods perform in general, it is
natural for the scope to be generic. However, results from a specific
study will be communicated more effectively when statisticians recognize
that the scope is specific for their audience. 

Scope is related to Cox's distinction between ``long-run behavior''
and a ``specific instance'' but differs in that the collection of
outcomes when the scope is generic need not be constructed in the
long-run. An interpretation for the confidence interval having generic
scope is given below that does not require repeated sampling. 

\subsection{Focus - Population or Model\label{subsec:Subject---Population}}

Scope applies to the interpretation of random phenomena whether or
not these are used for inference. Focus is meaningful only in the
context of statistical inference where we are concerned with an \emph{unknown}
distribution of numerical values. We call this distribution, whether
it be measurements on individuals in a population or values obtained
from random phenomena, the population distribution, or simply the
population when the context makes it clear that we are considering
a distribution of numerical values rather than a collection of individuals. 

Statistical inference proceeds by positing that a known distribution,
the model, is the same as, or an approximation to, the unknown population
distribution. While statistical inference is always concerned with
the population distribution, some inference procedures address the
population directly and others indirectly using one or more models
for the population. That is, the focus of an inference procedure can
be on the population or a model. 

The probability calculated for the clinical trial is a $p$-value
and the calculation of any $p$-value requires the specification of
a model (determined by the null hypothesis along with other assumptions).
Unless the population is the same as the model, it is difficult to
interpret the $p$-value as directly describing the population. 

On the other hand, probability used to describe confidence intervals
can have as its focus either the population or a family of models
for the population. For the former, the interpretation of a 95\% confidence
interval for the mean, say .03 to 41.83, is that this interval was
the result of an interval generating procedure applied to the population
that has the property that 95\% of the intervals from this procedure
contain the population mean. Since 95\% describes the procedure and
not the specific interval, the scope of this interpretation is generic
and the focus is the population. 

\citeauthor{fisher:1949} (1949, pages 190-191) provides the following
interpretation.
\begin{quote}
An alternative view of the matter is to consider that variation of
the unknown parameter, $\mu$, generates a continuum of hypotheses
each of which might be regarded as a null hypothesis, which the experiment
is capable of testing. In this case the data of the experiment, and
the test of significance based upon them, have divided this continuum
into two portions. One, a region in which $\mu$ lies between the
limits 0.03 and 41.83, is accepted by the test of significance, in
the sense that the values of $\mu$ within this region are not contradicted
by the data, at the level of significance chosen. The remainder of
the continuum, including all values of $\mu$ outside these limits,
is rejected by the test of significance.
\end{quote}
Here the focus is on a collection of models. The scope is specific
because each model is assessed in terms of how extreme the specific
data would be for that model. 

Simply checking whether a parameter value is in the interval shortchanges
the inferential value of the confidence interval. The endpoints serve
as guideposts indicating which models are such that the data would
be unlikely enough to elicit doubt regarding the model. For models
having mean slightly less than 0.03 the \emph{p-}value is slightly
less than 0.05 and for models having mean slightly greater than 0.03
the \emph{p-}value is slightly greater than 0.05. Similar comments
hold for models with means near 41.83. 

\section{Urn Models}

Urn models are a conceptual construction that provide a convenient
tool for describing inferential results in terms of classical probability.
One should conceive of a bowl filled with $N$ balls that are indistinguishable
in regard to their possible selection but completely distinguishable
in terms of at least one feature. This distinguishable feature is
needed to count the balls. The urn model is an example of a multiset
which is like a set except multiplicities are allowed. For sets, $\left\{ 1,2\right\} \cup\left\{ 2,3\right\} =\left\{ 1,2,3\right\} $
while for urns, $\lfloor1,2\rfloor\cup\lfloor2,3\rfloor=\lfloor1,2,2,3\rfloor$.
Unions and other basic set operations used below also hold for multisets. 

\subsection{Population Urn}

A population can be described using the conceptional construction
of an urn model. This model may be thought of as a bowl that contains
one ball for each member in the population. For a variable of interest
$X$, the population urn $\lfloor X\rfloor_{pop}$ is the bowl where
the numerical value for each member is written on the corresponding
ball. In most cases the values on the balls and the number of balls
$N$ are unknown. From the population urn we construct another urn
$\lfloor X\rfloor_{pop}^{n}$ containing ${N \choose n}$ balls. Each
sample of $n$ balls taken from $\lfloor X\rfloor_{pop}$ is represented
by one ball in $\lfloor X\rfloor_{pop}^{n}$; this ball is labeled
with an $n$-tuple of values obtained from the balls of the corresponding
sample from $\lfloor X\rfloor_{pop}$. The only restriction on $n$
is that it is a positive integer not greater than $N$. Notationally,
this conceptual construction is 
\[
\lfloor X\rfloor_{pop}\overset{C_{n}}{\longrightarrow}\lfloor X\rfloor_{pop}^{n}
\]
where the arrow indicates an enumeration of all possible samples of
$n$ balls so that the observed sample corresponds to a ball $\left(x\right)^{obs}$
in $\lfloor X\rfloor_{pop}^{n}$.\footnote{Sampling plans other than SRS would require a different enumeration. } 

\subsection{Model Urns}

For inference regarding the population, a model is posited for $\lfloor X\rfloor_{pop}$
and the urn for the model is written $\lfloor X\rfloor_{\theta}$
because often there will be a set of models indexed by a parameter
$\theta\in\Theta$. To assess how well $\lfloor X\rfloor_{\theta}$
approximates $\lfloor X\rfloor_{pop}$, the observed sample $\left(x\right)^{obs}$
is compared to the possible samples in the model, $\lfloor X\rfloor_{\theta}^{n}$,
where 
\begin{equation}
\lfloor X\rfloor_{\theta}\overset{C_{n}}{\longrightarrow}\lfloor X\rfloor_{\theta}^{n}.\label{eq:sampling_model}
\end{equation}
Unlike $\lfloor X\rfloor_{pop}^{n}$, the $n$-tuples on all balls
in $\lfloor X\rfloor_{\theta}^{n}$ are known.\footnote{The number of balls in model urn $\lfloor X\rfloor_{\theta}$ need
not equal the number in the population urn. The relevant features
are proportions rather than counts. }

The samples in $\lfloor X\rfloor_{\theta}^{n}$ are compared to the
observed sample using a test statistic $T_{\theta}$, a real valued
function on $\mathbb{R}^{n}$. The value of the observed test statistic
is $t_{\theta}^{obs}=T_{\theta}\left(x\right)^{obs}$. The plausibility
of a specific model $\lfloor X\rfloor_{\theta_{o}}$ as an approximation
to $\lfloor X\rfloor_{pop}$ is assessed by comparing $\left(x\right)^{obs}$
to the samples in $\lfloor X\rfloor_{\theta}^{n}$. Specifically,
by finding the \emph{proportion} of balls whose test statistic value
is greater than or equal to $t_{\theta_{o}}^{obs}$. This proportion
is written as 
\begin{equation}
\mbox{Pr}\lfloor T_{\theta_{o}}\ge t_{\theta_{o}}^{obs}\rfloor_{\theta_{o}}^{n}\label{eq:pval_significance1}
\end{equation}
where
\begin{align}
\mbox{Pr}\lfloor T & \ge t\rfloor_{\theta}^{n}=\frac{{|\left\{ b\in\lfloor X\rfloor_{\theta}^{n}:T(b)\ge t\right\} |}}{|\lfloor X\rfloor_{\theta}^{n}|}.\label{eq:pval_significance}
\end{align}
 No randomizations were used to construct the model urn $\lfloor X\rfloor_{\theta_{o}}^{n}$.
However, for the proportion in (\ref{eq:pval_significance1}) to be
meaningful as a probability, the observed sample must have been obtained
using a simple random sample (SRS) from the population. Given this
randomization, the proportion in (\ref{eq:pval_significance1}) is
the $p$-value for testing $H:\theta=\theta_{o}$ using the test statistic
$T_{\theta_{o}}$. 

The $(1-\alpha)100$\% confidence interval\footnote{This notation and interpretation allow generalizing to a confidence
region. } for $\theta$ obtained from $\left(x\right)^{obs}$ is found by allowing
$\theta_{o}$ in (\ref{eq:pval_significance1}) to range over all
possible values for $\theta$, 
\begin{equation}
C_{\left(x\right)^{obs}}^{\alpha}=\left\{ \theta:\mbox{Pr}\lfloor T_{\theta}\ge t_{\theta}^{obs}\rfloor_{\theta}^{n}\ge\alpha\right\} .\label{eq:CI_obs}
\end{equation}
The interval in (\ref{eq:CI_obs}) represents all the models, indexed
by $\theta$, for which the observed data would not be in the most
extreme $\alpha100$\% observations as measured by the ordering of
the test statistic $T_{\theta}$. Even though the confidence interval
$C_{\left(x\right)^{obs}}^{\alpha}$ involves many models there is
still only one randomization that is required \textendash{} the randomization
used to obtain the data from the population. 

The procedural interpretation of the confidence interval can be described
using an urn of confidence intervals 
\begin{equation}
\lfloor X\rfloor_{pop}^{n}\longleftrightarrow\lfloor C^{\alpha}\rfloor_{pop}^{n}\label{eq:CI_procedure}
\end{equation}
where the urn on the right is obtained by letting $\left(x\right)^{obs}$
in (\ref{eq:CI_obs}) range over all possible samples in the population.

\subsection{Compared to Repeated Sampling}

The sampling urns for the population and for models are constructed
using enumeration. In contrast, the limiting relative frequency interpretation
involves the conceptual construction of an infinite sequence where
each term in the sequence is obtained by a hypothetical random sample.
Notationally, 
\begin{equation}
\lfloor X\rfloor_{pop}\overset{SRS_{n}}{\longrightarrow}(x)_{1},(x)_{2},\ldots\label{eq:lim_rel_freq_SRS_hypothetical}
\end{equation}
where $(x)_{i}$ is the $n$-tuple obtained from the $i$th hypothetical
sample. Because these are random samples, another sequence
\begin{equation}
\lfloor X\rfloor_{pop}\overset{SRS_{n}}{\longrightarrow}(x)'_{1},(x)'_{2},\ldots\label{eq:lim_rel_freq_SRS_hypothetical-1}
\end{equation}
could be used. The sequences in (\ref{eq:lim_rel_freq_SRS_hypothetical})
and (\ref{eq:lim_rel_freq_SRS_hypothetical-1}) are different but
have the same limiting relative frequency.

The structure in random sampling that allows the calculation of probabilities
is represented in the limit of an infinite sequence whose order is
immaterial to describing this structure. In contrast, the enumeration
used to construct $\lfloor X\rfloor_{pop}^{n}$ imposes no artificial
ordering and describes the structure without infinite limits.

For models, limiting relative frequency could be described using a
conceptual construction where $\lfloor X\rfloor_{pop}$ is replaced
with $\lfloor X\rfloor_{\theta}$ in (\ref{eq:lim_rel_freq_SRS_hypothetical}).
While actual random samples from a model can be useful for calculations,
hypothetical random samples are not required for interpretation since
all samples are known. Furthermore, when hypothetical randomizations
are used to interpret model probabilities, probabilities that are
independent of the data, these can be confused with hypothetical randomizations
from the population that are intimately connected with the data. \footnote{Section \ref{subsec:Potential-Comparisons} provides an example.}

Random variables are used to model data and, if $X_{rv}$ is a random
variable\footnote{Common notation would be $X$ but we are using $X$ to represent a
finite collection of values.}, then the terminology suggests thinking of $X_{rv}$ as generating
a sequence of values through repeated randomization

\begin{equation}
X_{rv}\overset{SRS_{n}}{\longrightarrow}(x)_{1},(x)_{2},\ldots.\label{eq:lim_rel_freq_SRS_RV}
\end{equation}
We use the notation $\lfloor\cdot\rfloor$ to emphasize that the model
is an aggregate of values rather than a generator of infinite random
sequences. When the aggregate is finite, the distribution of $\lfloor X\rfloor_{\theta}$
is described by proportions having integer denominator. When the aggregate
is infinite, the distribution of $\lfloor X_{rv}\rfloor_{\theta}$
is described by the proportion of areas under a curve.\footnote{If $X_{rv}$ is continuous the curve is the probability density function.
If $X_{rv}$ is discrete the proportion of lengths would described
the distribution.} 

Neither the definition nor interpretation of a probability model requires
randomization. Both the definition and interpretation of frequentist
inference require randomization but this need not be imagined as belonging
to a hypothetical repetition of randomizations. The randomization
required is the one that produced the data that were obtained
\[
\lfloor X\rfloor_{pop}\overset{SRS_{n}}{\longrightarrow}(x)^{obs}.
\]
To recognize the importance of this randomization from the population,
models are described using (\ref{eq:sampling_model}) rather than
(\ref{eq:lim_rel_freq_SRS_RV}).

\section{Confidence Intervals}

The Fisher interpretation for the observed interval is naturally described
without repeated sampling using $C_{\left(x\right)^{obs}}^{\alpha}$.
The interpretation of a confidence interval as having been produced
by a procedure is typically described using repeated sampling. Section
\ref{subsec:CI_rv} shows that, in fact, a single random sample can
be used for the procedural interpretation. Section \ref{subsec:Complementary-Interpretations}
compares the single random sample interpretations of $C_{\left(x\right)^{obs}}^{\alpha}$
and $\lfloor C^{\alpha}\rfloor_{pop}^{n}$. 

\subsection{$\lfloor C^{\alpha}\rfloor_{pop}^{n}$\label{subsec:CI_rv}}

\citet{Greenland2016} provide the following interpretation for the
95\% confidence interval,
\begin{quotation}
$\ldots$ the 95\% refers only to how often 95\% confidence intervals
computed from very many studies would contain the true effect if all
the assumptions used to compute the intervals were correct.
\end{quotation}
It seems the word ``only'' is used to discourage other procedural
interpretations since earlier in their paper the observed confidence
interval is described in terms of testing which we understand to be
Fisher's interpretation. 

Even if the word ``only'' applies just to the procedural interpretation,
this statement is too strong. As the urn models show, this interpretation
need not be described in terms of limiting relative frequency. When
the family of models contains the true model, $\lfloor X\rfloor_{pop}=\lfloor X\rfloor_{\theta^{*}}$
for some $\theta^{*}$, then the urn $\lfloor C^{.05}\rfloor_{pop}^{n}$
defined by (\ref{eq:CI_procedure}) has the property that 95\% of
these intervals contain the true effect, $\theta^{*}$. The proportion
0.95 is a probability when each interval in $\lfloor C^{.05}\rfloor_{pop}^{n}$
is given an equally likely chance of being selected; i.e., the observed
data were obtained by an SRS from the population. The procedural interpretation
for the confidence interval does not require the procedure to be repeated
many times, just as understanding Cada's probability of winning did
not require repeatedly shuffling the remaining poker cards.

This requirement of conceptualizing very many studies leads to an
unnecessary criticism of a common (mis)interpretation regarding an
observed confidence interval:
\begin{quotation}
There is a 95\% chance that the population mean is between 0.03 and
41.83.
\end{quotation}
A standard response is ``Either the mean is between these values
or it is not. The values 0.03, 41.83, and the population are not random
so probability is not meaningful here.''\footnote{For a recent version of this response see \citet{Anderson2019TAS}. } 

This statement warrants caution rather than correction. To understand
how this can be a reasonable interpretation we consider a version
of the North Carolina Pick-3 Lottery where a Statistics professor
buys 1,000 Pick-3 tickets, one for each possible combination of three
digits from 000 to 999. The tickets are partitioned so that 20 tickets
are placed into each of 50 envelopes that are labeled with the names
of the 50 students in her class. The drawing is on Wednesday and at
Tuesday's lecture the professor asks Bob what is the probability that
his envelope has the winning ticket. Bob responds 1 in 50. The professor
will distribute the envelopes at Thursday's lecture. 

Before distributing the envelopes on Thursday, Bob is asked the same
question and again gives the probability of 1 in 50. Should the professor
correct Bob and say that either he has or has not won, and that probability
no longer applies? We think not. It is still meaningful to say the
probability for each student is 1 in 50. 

However, the situation on Tuesday is different from that on Thursday,
and recognizing this difference indicates the necessary caution. On
Thursday when the first envelope is opened the probability of the
remaining envelopes changes to 1 in 49 or to 0. If the envelopes had
been distributed before the drawing, any envelope could be opened
and the probability would remain 1 in 50. 

The interpretation ``There is a 95\% chance that the population mean
is between 0.03 and 41.83'' is incorrect when there is additional
information from the population (i.e., opened envelopes). In particular,
this interpretation cannot be used when there are two observed confidence
intervals from the same population \textendash{} let alone, ``very
many studies'' as the above repeated sampling interpretation requires.
However, without additional information from the population this statement
provides a reasonable description of the information in the data concerning
the population mean. \citeauthor{coxandv.hinkley1979} (1979, pages
227-228) also consider interpreting the observed interval in terms
of probability reasonable given appropriate cautions.

\subsection{Complementary Interpretations\label{subsec:Complementary-Interpretations}}

In terms of scope and focus the interpretations represented by $C_{\left(x\right)^{obs}}^{\alpha}$
and $\lfloor C^{\alpha}\rfloor_{pop}^{n}$ are very different. The
interval $C_{\left(x\right)^{obs}}^{\alpha}$is specific to the data
that was observed and the focus is on a collection of models. The
collection of intervals $\lfloor C^{\alpha}\rfloor_{pop}^{n}$ is
generic and the focus is on the population. 

The interpretations also differ in the assumptions that are required.
The urn $\lfloor C^{\alpha}\rfloor_{pop}^{n}$ cannot be constructed
directly since the population is unknown but relies on the assumption
that there is a model with parameter $\theta^{*}$ such that $\lfloor X\rfloor_{\theta^{*}}$
is a close approximation to $\lfloor X\rfloor_{pop}$. This assumption
is not required for the interpretation represented by $C_{\left(x\right)^{obs}}^{\alpha}$
.

Coverage probability and expected length apply to $\lfloor C^{\alpha}\rfloor_{pop}^{n}$
but not to $C_{\left(x\right)^{obs}}^{\alpha}$. When intervals are
defined with these two criteria in mind but without inverting a test,
there is great flexibility in how individual intervals are chosen.
As a result, observed intervals can have poor properties when interpreted
in terms of testing.\footnote{This issue arises when the sample space is discrete and the intervals
are considered too conservative in terms of coverage probability.
See, for example, \citet{Vos2008}.} To maintain fidelity to the Fisher interpretation, \citet{vos_hudson_TAS_2005}
introduce the criteria $p$-confidence and $p$-bias that apply to
$C_{\left(x\right)^{obs}}^{\alpha}$. 

\section{$P$-values\label{subsec:Potential-Comparisons}}

Confidence intervals allow for an interpretation that is population
focused. Interpreting $p$-values in terms of population focus can
lead to problems. As an example we consider the issue of potential
comparisons raised by \citet{gelman2016} who claims 
\begin{quote}
$\ldots$ to compute a valid \emph{p}-value you need to know what
analyses \emph{would have been done} had the data been different.
Even if the researchers only did a single analysis of the data at
hand, they well could\textquoteright ve done other analyses had the
data been different.
\end{quote}
\quad{}We cannot be certain of Gelman's interpretation for the \emph{p-}value
but\emph{ }the proportion in (\ref{eq:pval_significance1}) is a valid
\emph{p}-value and requires only a single random sample. Gelman considers
repeated sampling from the population but the \emph{p}-value is a
probability that describes a model, not the population. Comments by
\citeauthor{fisher1959statistical} (1959, page 44) apply here
\begin{quotation}
In general tests of significance are based on hypothetical probabilities
calculated from the null hypotheses. They do not generally lead to
any probability statements about the real world, but to a rational
and well-defined measure of reluctance to the acceptance of the hypotheses
they test.
\end{quotation}
\quad{}Certainly \emph{p}-values can be misused but Gelman's statement
is too strong because it makes $p$-values invalid even when there
has been no \emph{actual} misuse. A potential misuse of a $p$-value,
or any inference procedure, does not invalidate a single instance
of proper use. Consider the following example from Texas Hold'em Poker.
A gambler calculates the probability of making a specific hand based
on the proportion of unseen cards. This calculation is done under
the following conditions: he is well rested, sober, and knows the
dealer, and he has no reason to suspect cheating. The result of this
calculation is a valid probability. The gambler's wife might say that
if he were to play too much poker, then he would become sleepy, drink
too much, and gamble at shady establishments. Regarding the long run
outcome of his gambling, these are legitimate concerns that bring
the validity (utility) of future probability calculations into questions.
However, these potentialities do not affect the gambler's specific
calculation made under the actual conditions. The scope for the gambler
is specific while for his wife it is generic. 

The reader might find differences between our example and the discussion
of potential comparisons. Our hope is that we could agree that hypothetical
long run sampling is problematic when used to address a specific instance,
and our point is that repeated sampling is not required to interpret
inference for the data actually observed. 

\section{Power }

We have seen that confidence intervals and $p$-values can be interpreted
using a single random sample. Power calculations are done before data
have been collected and do not require any randomization or hypothetical
repetitions. This is in contrast to how power is often discussed.
For example, \citet{Greenland2016} describe power as a probability
``defined over repetitions of the same study design and so is a frequency
probability.''

Power calculations are done by comparing the model specified by a
null hypothesis to a competing model. The urn $\lfloor X\rfloor_{o}^{n}$
of the null model is compared to the urn $\lfloor X\rfloor_{1}^{n}$
of the competing model in terms of a test statistic $T_{o}$. Specifically,
the significance level $\alpha$ defines a value $t^{*}$ such that
\[
\mbox{Pr}\lfloor T_{o}\ge t^{*}\rfloor_{o}^{n}=\alpha
\]
and the power $\beta$ is given by
\[
\mbox{Pr}\lfloor T_{o}\ge t^{*}\rfloor_{1}^{n}=\beta.
\]

Both $\alpha$ and $\beta$ are proportions. The power is the proportion
of all samples of size $n$ from the competing model (posited as an
approximation to the population) that are more extreme than $t^{*}$.
Power calculations based on random variables are conducted in the
same way but now proportions with integer denominator are generalized
to proportions of area or a more general measure. These proportions
are meaningful as probabilities and useful for inference regarding
the population when the observed data is obtained by an actual randomization
from the population. Hypothetical repetitions from the population
or one of the models are not required.

\section{Discussion}

Describing the observed confidence interval as having been obtained
from a procedure is often the only interpretation that is considered,
but there are authors who recognize Fisher's interpretation. Examples
include, \citet{oscarkempthorne1971} who call Fisher's interpretation
a \emph{consonance interval} and \citet{deborahmayo2018} who describes
inference in terms of severe testing that appears to be very close
to Fisher's interpretation. 

Other authors also see pitfalls with the introduction of the concept
of infinity. For example, \citeauthor{hacking1976logic} (1976, p.
7) ``However much they have been a help, I shall argue that hypothetical
infinite populations only hinder full understanding of the very property
von Mises and Fisher did so much to elucidate.''

We have restricted urns to be finite for simplicity. Allowing an urn
to have an infinite number of balls results in a \emph{statistical
ensemble. }According to the Wikipedia entry (\citeyear{wiki:Statistical_Ensemble}) 
\begin{quote}
... an \textbf{ensemble} (also \textbf{statistical ensemble}) is an
idealization consisting of a large number of virtual copies (sometimes
infinitely many) of a system, considered all at once, each of which
represents a possible state that the real system might be in.
\end{quote}
A single simple random sample of $n$ individuals from a population
creates a statistical ensemble where the possible states consist exactly
of the possible samples of size $n$ from the population. 

The conceptualization of a statistical ensemble differs from repeated
sampling in that a large number is considered \emph{all at once} and
this idea avoids several pitfalls associated with repeated sampling.
Repeated sampling and terms such as ``long run'' introduce the notion
of time even though time is not included in the definition of probability.
Adding to the confusion is that when the scope is generic, such as
a statistician defining procedures in terms of ``how they would perform
were they used repeatedly'', time fits naturally in that particular
interpretation. Furthermore, repetition generates a sequence and the
order of this sequence has nothing to do with the structure of the
collection so the idea of independence is needed to appropriately
describe a random sequence. By considering the collection all at once,
whether it is balls in an urn or states of an ensemble, these complications
are avoided. A statistical ensemble can be applied when the scope
is generic or specific but is especially useful in the latter case. 

Recognizing that the focus can be either the population or the model
sheds light on the role of randomization in statistical inference.
Randomly selecting data from a population is fundamental for making
inferences about the population, and models are used to make inferences,
but no randomizations from the model are required. Hypothetical repeated
randomizations may be introduced as a means to interpret the probability
obtained from the model, but these hypothetical randomizations, and
the consequent confusion with the required randomization from the
population, can be avoided by using urn models or statistical ensembles. 

\bibliographystyle{Chicago}
\bibliography{vos}
 
\end{document}